\documentclass{article}

\usepackage{PRIMEarxiv}

\usepackage[utf8]{inputenc} 
\usepackage[T1]{fontenc}    
\usepackage{hyperref}       
\usepackage{url}            
\usepackage{booktabs}       
\usepackage{amsfonts}       
\usepackage{nicefrac}       
\usepackage{microtype}      
\usepackage{lipsum}
\usepackage{fancyhdr}       
\usepackage{graphicx}       
\graphicspath{{media/}}     
\usepackage{subcaption}
\usepackage[section]{placeins}

\title{Exploring Global Climate Cooperation through AI: An Assessment of the AI4GCC Framework by simulations}

\author{
  Xavier Marjou, Arnaud Braud, Gaël Fromentoux \\
  Orange Innovation \\
  Orange \\
  Lannion\\
  \texttt{\{xavier.marjou, arnaud.braud, gael.fromentoux\}@orange.com} \\
}

\begin{document}
\maketitle

\begin{abstract}
In scenarios where a single player cannot control other players, cooperative AI is a recent technology that takes advantage of deep learning to assess whether cooperation might occur. One main difficulty of this approach is that it requires a certain level of consensus on the protocol (actions and rules), at least from a majority of players. In our work, we study the simulations performed on the cooperative AI tool proposed in the context of AI for Global Climate Cooperation (AI4GCC) competition. We experimented simulations with and without the AI4GCC default negotiation, including with regions configured slightly differently in terms of labor and/or technology growth. These first results showed that the AI4GCC framework offers a promising cooperative framework to experiment with global warming mitigation. We also propose future work to strengthen this framework.
\end{abstract}


\section{Introduction}

Climate and resource issues generate structural transformations, sometimes brutal and unpredictable, in the business environment of all economic actors. As a result, they become major points of tension. In this context, many agencies and companies are studying the new environmental, social and corporate governance (ESG) reality to recommend new ESG-sensitive policies and investments, leading to many new initiatives.

To meet this need for anticipation and evolution of strategic thinking, there is a need for prospective scenarios, as well as new tools to virtually experiment them, as highlighted by the Carbon 4 IRIS initiative\footnote{e.g. \url{https://www.carbone4.com/lancement-iris-initiative}.}. However, given that environmental and social issues generally involve multiple players, it is not always possible for rule-makers to control all players. As a consequence, there is a need to design policies motivating cooperative solutions. For instance \cite{9830283}, \cite{legleau:tel-03813640} proposed and evaluated multiple cooperation scenarios allowing Mobile Network Operators (MNOs) to cooperate to save energy during low-activity hours.

Regarding climate warming, AI for global climate cooperation (AI4GCC) \cite{https://doi.org/10.48550/arxiv.2208.07004} is a new community initiative that recently launched an interdisciplinary challenge to identify the most suited negotiation protocol to reach the best compromise between economic and climatic concerns. To frame this work, they provided a tool under the form of a MARL environment that allows multiple simulated regions to interact in order to collectively mitigate climate change. 

In this paper, we experimented the AI4GCC tool with two configurations: one with agents who do not negotiate, another one with agents who negotiate using the default negotiation protocol implemented by the framework. We observed that this negotiation allows to decrease the global temperature. In addition, we also highlighted some current limitations of the framework.

\section{AI4GCC Framework}

\subsection{Model overview}

The AI4GCC framework \cite{https://doi.org/10.48550/arxiv.2208.07004} comes under the form of a reinforcement-learning (RL) \cite{sutton1998introduction} environment that allows instantiating multiple agents, each representing a region. They interact during an episode of T=20 iteration steps, either with or without negotiation. Each step represents a 5-year period ($\Delta$), hence a full episode represents a duration of 100 years. At each step $t$, the RL environment returns a distinct step-reward ($u_i,_t$) to each RL agent (aka region $i$), which is proportional to the labor $l$ and the consumption $c$ of the region. At the end of each episode, the sum of the step-rewards provides a \emph{regional episode-reward} ($u_i$). During training, each agent periodically calculates its regional episode-reward ($u_i$) (mean value over the last 100 episodes) and the framework also derives a \emph{collective episode-reward} $u$ (which is also known as \emph{episode\_reward\_mean} in RLlib Ray software).

By default, the tool comes with N=27 pre-configured regions as shown in Table ~\ref{tab:table-config}.

\begin{equation}
\begin{array}{ccc} 
u_i,_t = \frac{(l_i,_t / 1000.0) * (\left(\frac{c_i,_t}{(l_i,_t / 1000.0)} + \epsilon\right)^{1 - \alpha} - 1)}{1 - \alpha} , &
u_i = \sum_{t=1}^{T} u_i,_t
\label{eqn:region_episode_reward} , &
u = \sum_{i=1}^{N} u_i \\
\end{array}
\label{eqn:utility}
\end{equation}

Two main updates happen between two consecutive timesteps:
\begin{itemize}
\item The labor ($l$) update is based on a \emph{long-term population} size ($l_{a;i}$) and a \emph{convergence speed} value ($l_{g;i}$).
\item The consumption ($c$) is based on a multiple parameters, including a delta\_A
 parameter ($\delta_a$).
\end{itemize}

\begin{equation}
l_i,_{t+1} = l_{i,t} \times \left(\frac{1 + l_{a;i}}{1 + l_{i,t}}\right)^{l_{g;i}}
\label{eqn:labor_upgrade}
\end{equation}

\begin{equation}
c_i,_{t+1} = \left( dom\_pref \cdot c_{dom{_i,_{t+1}}}^{sub\_rate} + \sum_{i=1}^{N} for\_pref_i \cdot c_{for_i}^{sub\_rate} \right)^{\frac{1}{sub\_rate}}
\end{equation}
with:
\begin{equation}
c_{dom{_i,_{t+1}}}  = max(0, gross\_output - investment - \sum_{r=1}^{N} exports_i)
\end{equation}

\begin{equation}
gross\_output_{i,t+1}  = damages_{t+1}  \times (1 - abatement\_cost_{i,t+1} ) \times production_{i,t+1} 
\end{equation}

\begin{equation}
damages,_{t+1}  = \frac{1}{1 + a_1 \cdot atm\_temp_{t+1} + a_2 \cdot atm\_temp_{t+1}^{a_3}}
\end{equation}

\begin{equation}
production_i,_{t+1} = technology_i,_{t+1} \times capital_{i,t+1}^{\gamma} \times \left(\frac{l_i,_{t+1}}{1000}\right)^{1-\gamma}
\end{equation}

\begin{equation}
capital_i,_{t+1} = capital\_depreciation_i,_{t+1} \times capital_i,_{t} + \Delta \times (savings \times gross\_output)
\end{equation}

\begin{equation}
technology_i,_{t+1} = (e^{0.0033} + g_{a,i} \times e^{-\delta_{a,i} \times \Delta \times (t - 1)}) \times technology_i,_{t}
\label{eqn:production}
\end{equation}

Among these equations and parameters, the \emph{labor} (a.k.a. population) in equation \ref{eqn:labor_upgrade}) and \emph{technology} (a.k.a. total factor productivity) in equation \ref{eqn:production} are two important levers that we want vary in order to assess the reaction of the framework.

Note: equations 1 to 9 are an excerpt from the equations in the AI4GCC white-paper \cite{https://doi.org/10.48550/arxiv.2208.07004}. Please refer to this white-paper for more details.

\subsection{Framework Limitations}

Although the current AI4GCC framework is already considerable and challenging to understand in its smallest economic details, we noticed three aspects that we could not easily implement and integrate into our experimentation.
\begin{itemize}
\item Configuring strong discontinuities like a sudden drop of labour (eg: due to a natural disaster), or a technology leap-forward (e.g. due to so-called AGI tools). It would be important to allow such hypotheses to be tested, but this would probably question the economic model.
\item Integrating cooperation metrics such the incentive-to-cooperate, safety and fairness (cf definition in \cite{legleau:tel-03813640}, as well as their use in a telecom use-case in \cite{9830283}). This would  require a framework allowing each agent to implement a specific regional policy.
\item Simulating more than 27 regions (e.g. 100-200 regions) in a reasonable training time. Hopefully, the results of the AI4GCC challenge will bring solutions like notion of clubs that will make this achievable.
\end{itemize}

\section{Experiment}

We used the RL software environment offered by AI4GCC\footnote{https://github.com/mila-iqia/climate-cooperation-competition}, in which we performed two modifications:
\begin{itemize}
\item In order to log the metrics associated to each region, we added callbacks in the environment. 
\item In order to respectively modify the configuration of \emph{labor} and \emph{technology} (cf. experiment 2), we modified the default AI4GCC values $l_{a;i}$ and $g_{a,i}$ of a region.
\end{itemize}
We used Ray 1.0.0 on an NVIDIA 2080-Ti GPU to perform our MARL experiments. Each test was performed from scratch 5 times (i.e.: one full training before each test) to estimate representative values for the mean and standard deviation.

\subsection{Experiment-1}
In a first stage, we used the default AI4GCC configuration of Table ~\ref{tab:table-config} for all agents (a.k.a. regions).
In order to estimate whether the negotiation protocol affects the economical ranking of regions, we performed two tests; one without negotiation and another one with negotiation. 
\begin{itemize}
\item Test-1-no-nego: evaluation without negotiation (no agent tries to negotiate)
\item Test-1-nego: evaluation with negotiation (all agents try to negotiate, based on the default negotiation).
\end{itemize}
Each test was performed after training from scratch on $40,000$ episodes.

\subsection{Experiment-2}

In a second stage, we experimented different configurations than the default AI4GCC configuration of Table ~\ref{tab:table-config} either for all 27 regions, or for only one region (15, 19, or 6). Similarly to experiment-1, we carried out one test with negotiation and another one without negotiation, hence $4*2 = 8$ tests.

To input different \emph{labor and technology configurations} (LTCs) in the AI4GCC environment, we extracted the default AI4GCC $l_{a;i}$ and $g_{a,i}$ values and modified them to vary the labor and/or technology values respectively. Since early experiments showed that the current framework can not withstand large configuration disruptions such as a drastic drop of population in the long run or at one point, we remained relatively conservative and only experimented the following variations: $[10\%, 0\%, 10\%]$ for $l_{a;i}$ and $[-10\%, 0\%, 10\%]$ for $g_{a,i}$, which led to a set of $3*3=9$ LTCs, i.e. 9 subtests per test. 

The list of eight tests is as follows. 
\begin{itemize}
\item Evaluation without negotiation
    \begin{itemize}
    \item Test-2-a-no-nego: all agents have the same LTC.
    \item Test-2-h-no-nego: only a high-ranked region (region 15) has a different LTC.
    \item Test-2-m-no-nego: anly a middle-ranked region (region 19) has a different LTC.
    \item Test-2-l-no-nego: only a low-ranked region (region 6) has a different LTC.
    \end{itemize}
\item Evaluation with negotiation
    \begin{itemize}
    \item Test-2-a-nego: all agents have the same LTC.
    \item Test-2-h-nego: only a high-ranked region (region 15) has a different LTC.
    \item Test-2-m-nego: only a middle-ranked region (region 19) has a different LTC.
    \item Test-2-l-nego: only low-ranked region (region 6) has a different LTC.
    \end{itemize}
\end{itemize}

Since early experiments showed that training for $10,000$ episodes did not significantly change the results compared to training for $40,000$ episodes, we performed each subtest after training from scratch for only $10,000$ episodes in order to save energy (in kWh).

\section{Results}

\subsection{Experiment-1}

Table~\ref{tab:expe-1-results} shows that the global temperature increase is less important at the end of the episode than without negotiating, which is a nice and major result, but which comes with a lower collective reward.

Table~\ref{tab:table-expe-1} goes deeper into the details by describing the results per region. The values are based on the mean value on the last 100 episodes of the training (i.e. once the model has finished its training).  Although each region lost between 0 to 20\% of its utility, the results showed that the negotiation protocol led to minor modifications regarding the ranking of each region, which is an advantage to incentive regions to participate in the framework. 

In addition, Table \ref{tab:action} summarizes the mean values for each of the five default actions across the 27 regions both for the negotiated mode and non-negotiated mode simulations. It shows that the default negotiation leads to higher mitigation rates, lower saving rates and, surprisingly, to maximum exports close to zero. Although we could not explain the root cause, it nevertheless suggests that the negotiation step is useful as it induces agents to perform slightly different actions, which leads to a lower global temperature than in the baseline scenario (no negotiation).

\begin{table}
\begin{center}
\begin{tabular}{|c|c|c|}
  \hline
  test & global temperature increase & collective episode-reward \\
  \hline
  test-1-no-nego & $5.8\pm0.2$ & $165.5\pm0.5$ \\
  test-1-nego & $3.0\pm0.3$     & $152.9\pm3.4$ \\
  \hline
\end{tabular}
\caption{Experiment-1's main results}
\label{tab:expe-1-results}
\end{center}
\end{table}

\begin{table*}[!b]
\begin{center}
    \begin{tabular}{|c|cc|cc|c|c|}
    \hline
    region id & $u_i$ no-nego & rank no-nego & $u_i$ nego & rank nego & gain & rank delta \\
    \hline
    0 & 5.8 & 11 & 5.0 & 11 & 0.86 & 0 \\
    1 & 2.2 & 19 & 2.0 & 20 & 0.90 & 1 \\
    2 & 1.4 & 22 & 1.2 & 22 & 0.85 & 0 \\
    3 & 3.6 & 17 & 2.8 & 17 & 0.77 & 0 \\
    4 & 0.9 & 25 & 0.9 & 25 & 1.00 & 0 \\
    5 & 10.9 & 4 & 10.1 & 3 & 0.92 & -1 \\
    6 & 0.7 & 26 & 0.6 & 26 & 0.85 & 0 \\
    7 & 1.1 & 24 & 0.9 & 24 & 0.81 & 0 \\
    8 & 9.7 & 5 & 8.7 & 5 & 0.89 & 0 \\
    9 & 1.3 & 23 & 1.2 & 21 & 0.92 & -2 \\
    10 & 3.6 & 16 & 3.0 & 16 & 0.83 & 0 \\
    11 & 3.8 & 14 & 3.6 & 14 & 0.94 & 0 \\
    12 & 6.0 & 10 & 5.9 & 10 & 0.98 & 0 \\
    13 & 3.8 & 15 & 3.8 & 13 & 1.00 & -2 \\
    14 & 2.1 & 20 & 2.0 & 19 & 0.95 & -1 \\
    15 & 22.6 & 0 & 21.2 & 0 & 0.93 & 0 \\
    16 & 14.5 & 2 & 13.5 & 2 & 0.93 & 0 \\
    17 & 7.3 & 8 & 6.7 & 8 & 0.91 & 0 \\
    18 & 11.5 & 3 & 9.5 & 4 & 0.82 & 1 \\
    19 & 5.5 & 12 & 4.4 & 12 & 0.80 & 0 \\
    20 & 16.2 & 1 & 15.7 & 1 & 0.96 & 0 \\
    21 & 3.9 & 13 & 3.4 & 15 & 0.87 & 2 \\
    22 & 8.8 & 7 & 8.0 & 7 & 0.90 & 0 \\
    23 & 1.5 & 21 & 1.1 & 23 & 0.73 & 2 \\
    24 & 6.6 & 9 & 6.4 & 9 & 0.96 & 0 \\
    25 & 2.5 & 18 & 2.2 & 18 & 0.88 & 0 \\
    26 & 9.3 & 6 & 8.0 & 6 & 0.86 & 0 \\
    \hline
    $u$ & 165.5 &  & 150.5 &  &  &  \\    
    \hline
  \end{tabular}
  \caption{regional episode-rewards and collective episode-reward with and without negotiation}
  \label{tab:table-expe-1}
  \end{center}
\end{table*}

\begin{table*}[!b]
\begin{center}
\begin{tabular}{|c|c|c|}
  \hline
  action & test-1-no-nego & test-1-nego \\
  \hline
  mitigation rate & $0.009\pm0.005$ & $0.041\pm0.002$ \\
  saving rate & $0.011\pm0.002$     & $0.059\pm0.002$ \\
  max export & $1727\pm730$    & $1\pm3$ \\
  mean imports & $0.022\pm0.001$    & $0.024\pm0.002$\\
  mean tariffs & $0.021\pm0.001$    & $0.022\pm0.003$\\
  \hline
\end{tabular}
\caption{Actions values}
\label{tab:action}
\end{center}
\end{table*}

\begin{table}
\begin{center}
\begin{tabular}{|c|c|c|c|}
  \hline
  region & $u_i$  no-nego & $u_i$  nego & difference \\
  \hline
  15 & $22.3$ & $21.4$ & $-4\%$\\
  19 & $5.3$ & $4.6$ & $-14\%$\\
  6 & $0.6$ & $0.6$ & $-5\%$\\
  all & $6.0$ & $5.7$ & $-6\%$\\
  \hline
\end{tabular}
\caption{Average regional episode-rewards with and without negotiations}
\label{tab:deltas}
\end{center}
\end{table}

\subsection{Experiment-2}

The results (cf Figure 1 to Figure 8 in Annex) showed that conservative modifications of labor and/or technology configurations led to three observations:
\begin{itemize}
\item The tension between utility and temperature is confirmed.
\item Regardless of the LTC of each of the nine subtests, each region obtained a lower gain with negotiation compared to no-negotiation. However, the percentage of loss depends on the region (e.g. between $-14\%$ and $-4\%$ as indicated in Table~\ref{tab:deltas}), which might be a source of tension.
\item The ranking of the three tested regions (high-ranked, mid-ranked, or low-ranked region) did not change, regardless of the configured LTC.
\end{itemize}

Moreover, the results showed that a single region, regardless of its rank, could not have a significant impact on the collective outcome, both in the scenarios with and without negotiations.

\section{Discussion}
In its current form, the framework naturally encourages cooperation as every country suffers the same penalties from the damage function but also have no way of having a greedier approach to mitigating climate damage than investing in carbon reduction. We would encourage having regions have a potentially more selfish approach in order to capture reality a bit more by at least:
\begin{itemize}
    \item Having a different damage function per region, since global temperature doesn't impact regions the same way, an local temperature does not impact regions the same way either.
    \item Having a new course of action for every region called resiliency investments which only effect would be to reduce the impact of climate damage done on the economy at a regional scale. This could encompass the risk of badly designed negotiation protocols by allowing a less efficient selfish path that doesn't require any trust.
\end{itemize}
Another interesting evolution path would be to exploit the dynamism of the framework to use less continuous functions both for reward and damage by introduction a form a catastrophic event probability as the temperature rises and the social welfare of some region fails to catch up. This would however have negative consequences on both the explain-ability of the model and the training costs (as way more games would need to be played to capture those probabilities).

Finally it would be interesting to re-use the lesson learnt from this first approach, both from an AI perspective and from how we can use the results of RL AI to drive policy, with another modelling framework. It will also be challenging to see how frameworks that are going to be used by AIs differ from framework to be used by humans. Our two initial targets will be to use such models in a more telco oriented scenario and in a more ressource driven cross sectorial scenario.

\section{Conclusion}
We experimented the AI4GCC framework, which is a decision support tool using cooperative AI to simulate cooperative scenarios to mitigate global warming while preserving economic activities. Our experiments confirmed that the current default negotiation leads to mitigating global warming but the root cause incentivizing agents to lower economical reward remains to be identified. We also suggested to allow for a damage function more specific per region. As other initiatives aimed at mitigating environmental problems multiply, it will be interesting to study how they could reuse the AI4GCC framework.

\section*{Acknowledgments}
This work reused the AI4GCC code from MILA and Salesforce.

\bibliographystyle{unsrt}  
\bibliography{references}  

\section{Annex}

\begin{figure}
  \centering
  \includegraphics[width=\linewidth]{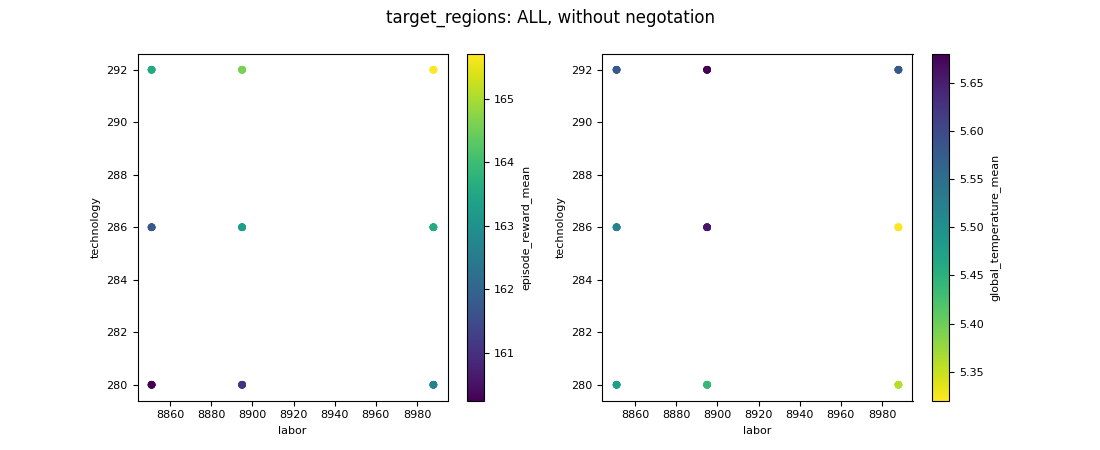}
  \caption{Test-2-a-no-nego results}
  \label{fig:res27_0}
\end{figure}

\begin{figure}
  \centering
  \includegraphics[width=\linewidth]{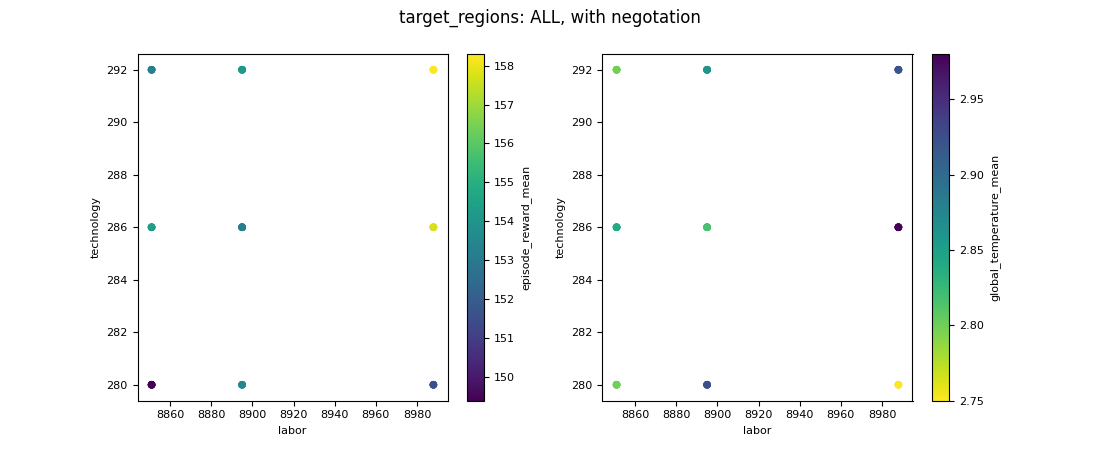}
  \caption{Test-2-a-nego results}
  \label{fig:res27_1}
\end{figure}

\begin{figure}
  \centering
  \includegraphics[width=\linewidth]{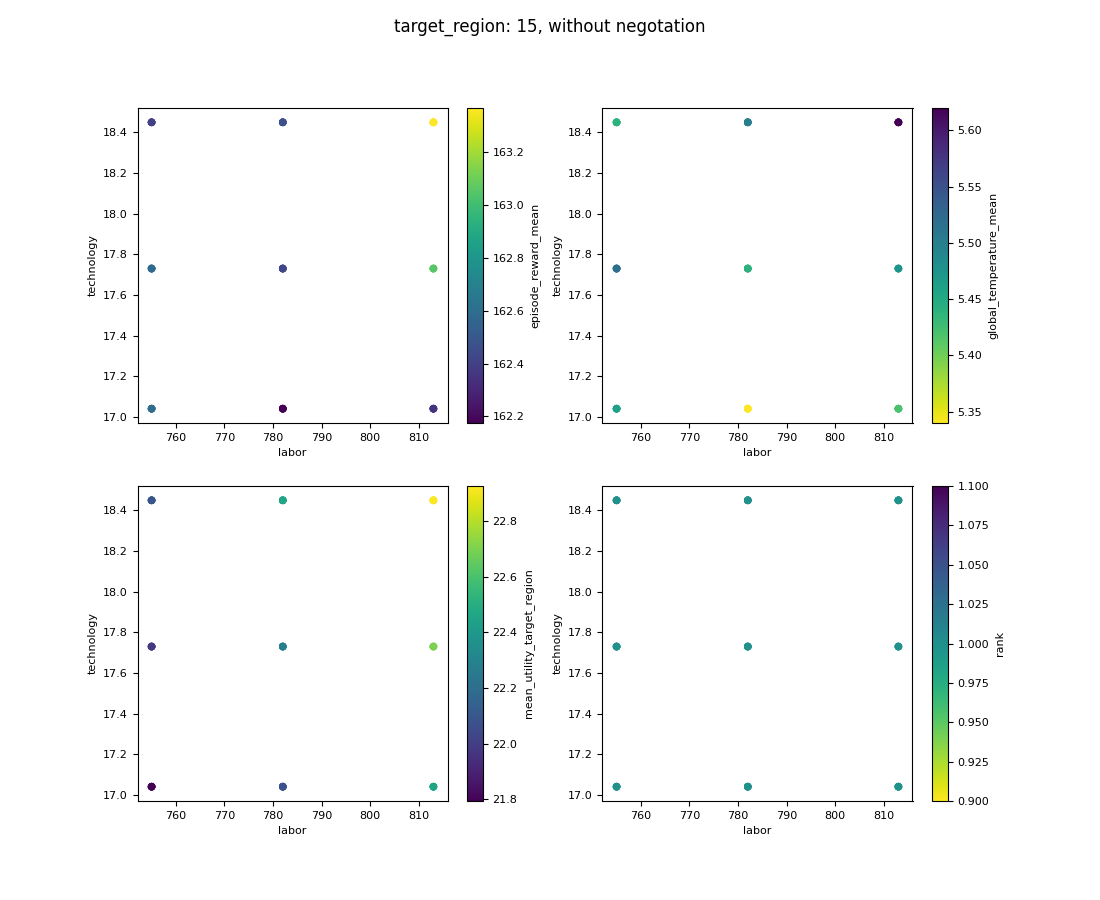}
  \caption{Test-2-h-no-nego results}
  \label{fig:res15_0}
\end{figure}

\begin{figure}
  \centering
  \includegraphics[width=\linewidth]{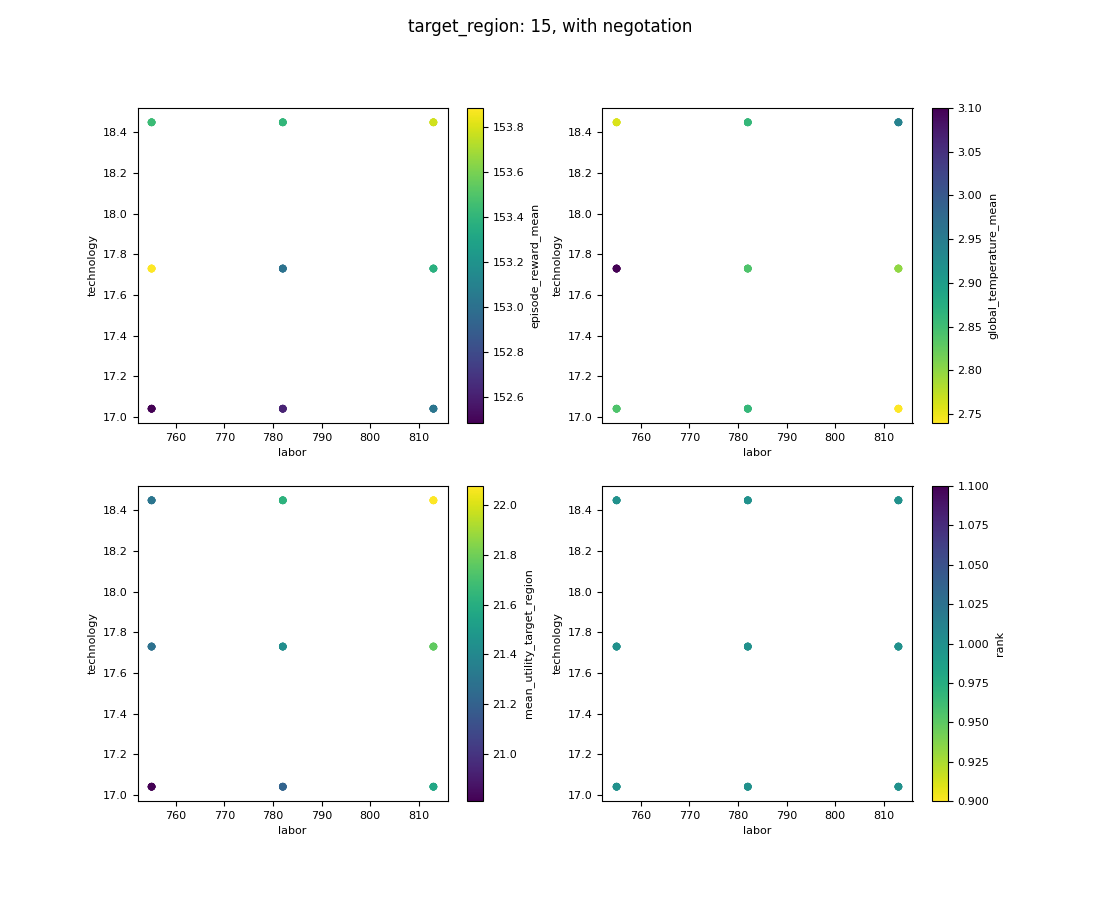}
  \caption{Test-2-h-nego results}
  \label{fig:res15_1}
\end{figure}

\begin{figure}
  \centering
  \includegraphics[width=\linewidth]{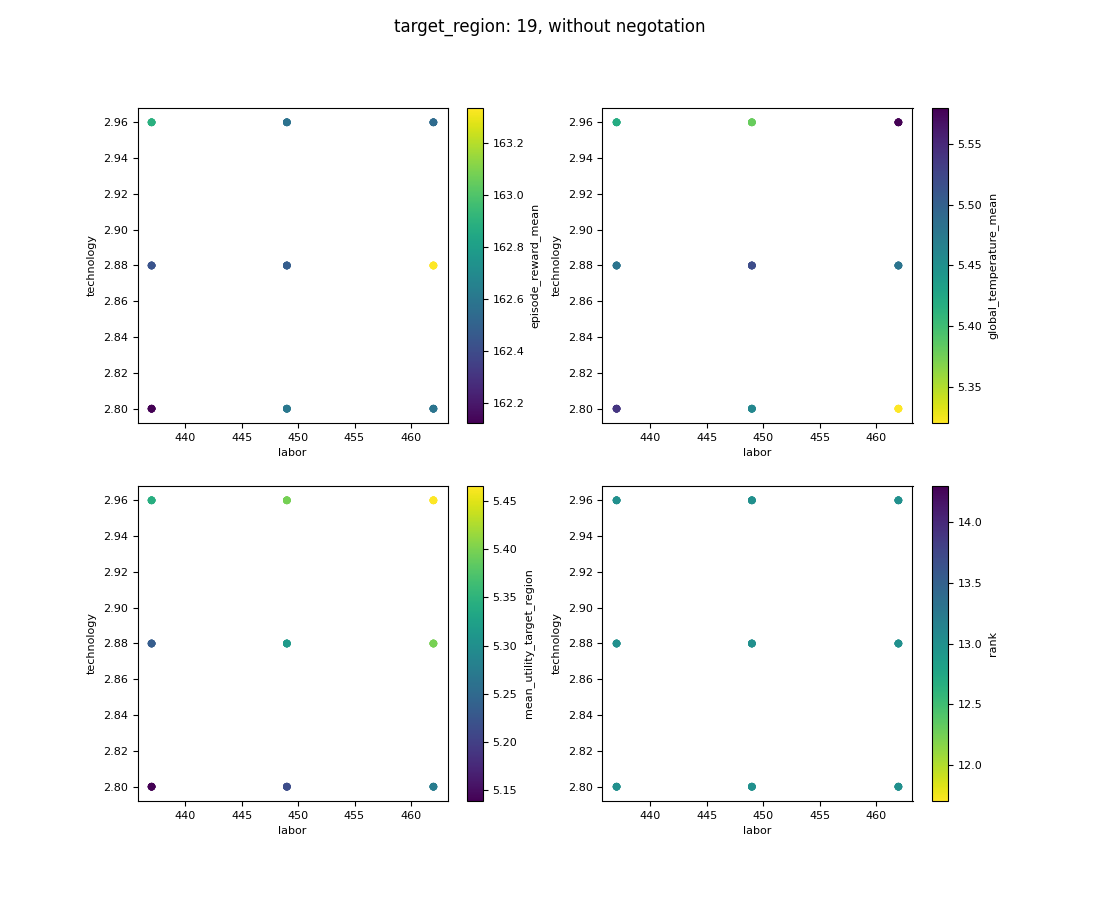}
  \caption{Test-2-m-no-nego results}
  \label{fig:res19_0}
\end{figure}

\begin{figure}
  \centering
  \includegraphics[width=\linewidth]{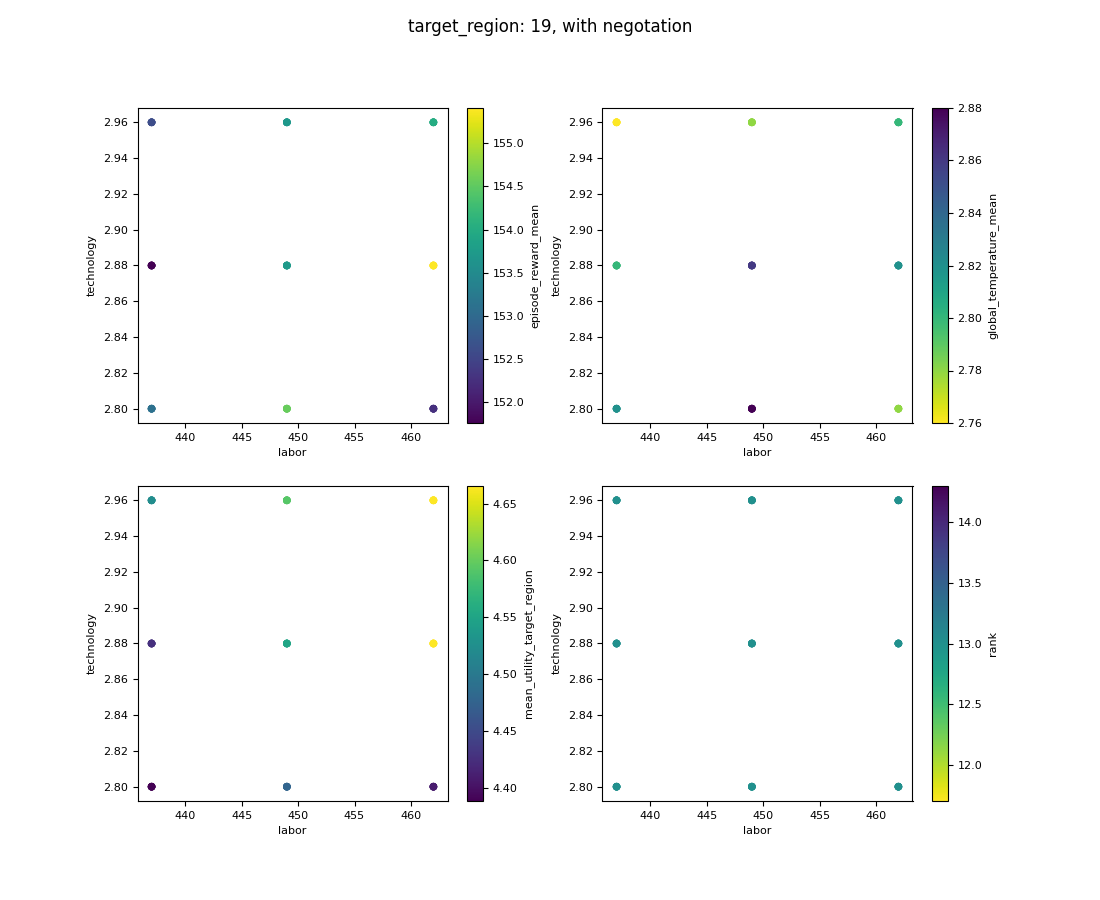}
  \caption{Test-2-m-nego results}
  \label{fig:res19_1}
\end{figure}

\begin{figure}
  \centering
  \includegraphics[width=\linewidth]{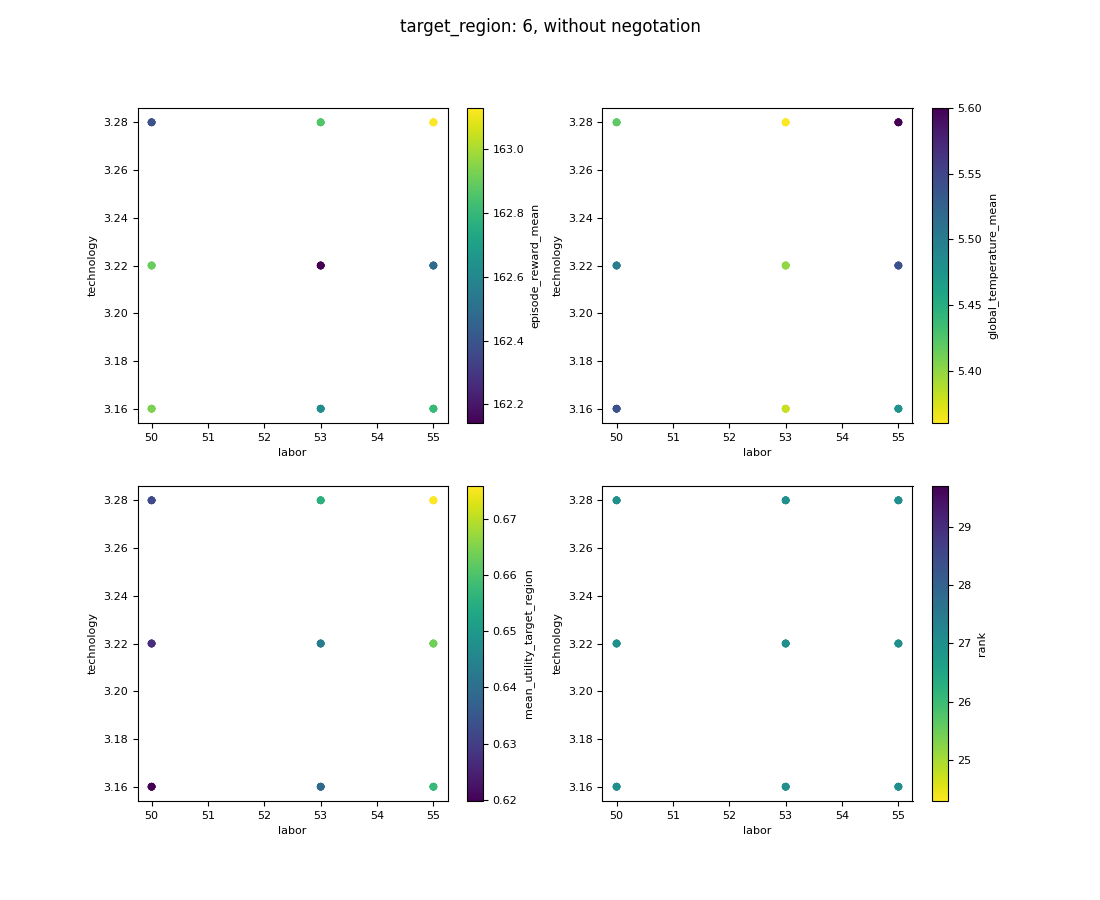}
  \caption{Test-2-l-no-nego results}
  \label{fig:res6_0}
\end{figure}

\begin{figure}
  \centering
  \includegraphics[width=\linewidth]{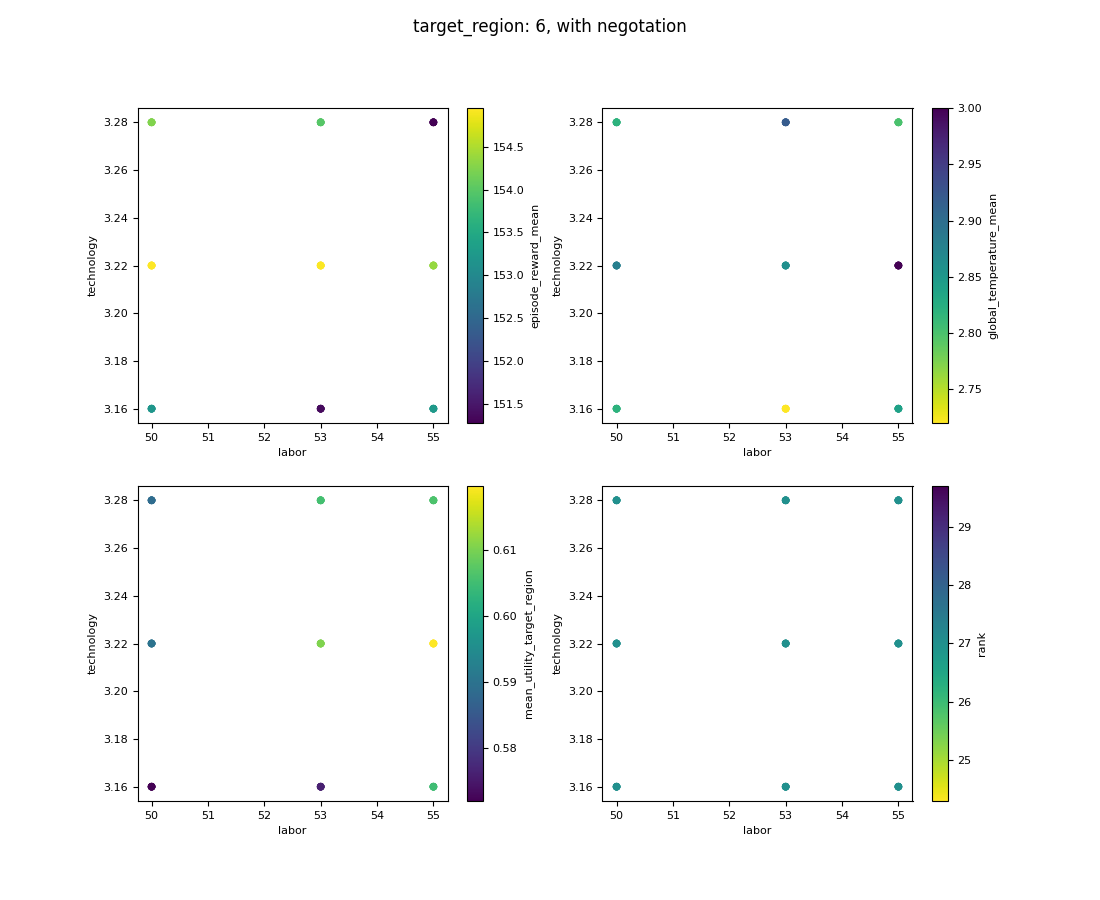}
  \caption{Test-2-l-no-nego results}
  \label{fig:res6_1}
\end{figure}

\begin{table}
    \begin{tabular}{|c|cccccccc|}
    \hline
    region\_id &    xA\_0 &    xK\_0 &     xL\_0 &      xL\_a &  xdelta\_A &   xg\_A &   xl\_g &  xsigma\_0 \\
    \hline
    0         &   1.872 &   0.239 &  476.878 &   669.594 &     0.139 &  0.122 &  0.034 &     0.456 \\
    1         &   8.405 &   3.304 &   68.395 &    93.497 &     0.188 &  0.103 &  0.058 &     0.529 \\
    2         &   3.558 &   0.109 &   64.122 &   135.074 &     0.161 &  0.127 &  0.026 &     0.816 \\
    3         &   1.927 &   1.424 &  284.699 &   465.308 &     0.244 &  0.134 &  0.024 &     1.221 \\
    4         &   8.111 &   0.268 &   28.141 &    23.574 &     0.163 &  0.106 & -0.057 &     0.290 \\
    5         &   4.217 &   3.184 &  548.754 &   560.054 &     0.170 &  0.095 &  0.080 &     0.302 \\
    6         &   2.491 &   0.044 &   46.489 &    59.988 &     0.058 &  0.049 &  0.037 &     0.420 \\
    7         &   2.525 &   1.080 &   69.194 &   100.016 &     0.346 &  0.079 &  0.029 &     1.010 \\
    8         &   2.460 &   0.184 &  513.737 &  1867.771 &     1.839 &  0.462 &  0.017 &     0.310 \\
    9         &  12.158 &   2.642 &   38.101 &    56.990 &     0.131 &  0.063 &  0.020 &     0.350 \\
    10        &   0.993 &   0.160 &  522.482 &  1830.325 &     0.086 &  0.065 &  0.019 &     0.235 \\
    11        &   5.000 &   2.289 &  165.293 &   230.191 &     0.183 &  0.071 &  0.027 &     0.419 \\
    12        &  29.854 &   2.020 &  165.751 &   216.927 &     0.088 &  0.075 & -0.002 &     0.254 \\
    13        &  23.315 &   3.039 &  109.395 &   143.172 &     0.088 &  0.075 & -0.002 &     0.254 \\
    14        &  29.854 &   0.687 &   56.355 &    73.755 &     0.088 &  0.075 & -0.002 &     0.254 \\
    15        &  10.922 &   0.606 &  705.465 &   532.497 &     0.096 &  0.168 & -0.016 &     0.781 \\
    16        &   9.634 &   0.608 &  465.607 &   351.448 &     0.096 &  0.168 & -0.016 &     0.781 \\
    17        &   8.621 &   0.453 &  239.858 &   181.049 &     0.096 &  0.168 & -0.016 &     0.781 \\
    18        &   3.190 &   0.129 &  690.002 &   723.513 &     0.054 &  0.068 & -0.013 &     0.949 \\
    19        &   2.034 &   0.381 &  455.401 &   477.518 &     0.054 &  0.068 & -0.013 &     0.949 \\
    20        &  13.220 &  16.295 &  502.410 &   445.861 &     0.252 &  0.074 & -0.033 &     0.170 \\
    21        &   3.190 &   0.044 &  234.601 &   245.994 &     0.054 &  0.068 & -0.013 &     0.949 \\
    22        &   6.387 &   1.094 &  317.880 &   287.533 &     0.194 &  0.237 & -0.053 &     0.840 \\
    23        &   2.481 &   0.090 &   94.484 &   102.997 &     0.203 &  0.201 &  0.037 &     1.665 \\
    24        &  10.853 &  17.554 &  222.891 &   168.351 &     0.005 & -0.000 & -0.012 &     0.285 \\
    25        &   4.135 &   1.002 &  103.294 &    87.418 &     0.158 &  0.123 & -0.063 &     0.601 \\
    26        &   2.716 &   1.034 &  573.818 &   681.210 &     0.097 &  0.101 &  0.043 &     0.638 \\
    \hline
  \end{tabular}
  \caption{AI4CC configuration parameters of the 27 regions.}
  \label{tab:table-config}
\end{table}

\end{document}